\documentclass{aastex}

\usepackage{apjfonts}
\usepackage{myemulateapj5}

\newcommand{\be}{  \begin{eqnarray} }
\newcommand{\ee}{  \end{eqnarray} }

\def\spose#1{\hbox to 0pt{#1\hss}}
\def\lta{\mathrel{\spose{\lower 3pt\hbox{$\mathchar"218$}}
     \raise 2.0pt\hbox{$\mathchar"13C$}}}
\def\gta{\mathrel{\spose{\lower 3pt\hbox{$\mathchar"218$}}
     \raise 2.0pt\hbox{$\mathchar"13E$}}}
\shorttitle{Turbulent Comptonization in Accretion Disks}
\begin{document}
\title{Turbulent Comptonization in Black Hole Accretion Disks}
\author{Aristotle Socrates, Shane W. Davis, and Omer Blaes}
\affil{Department of Physics, University of California, Santa Barbara,
CA 93106}

\begin{abstract}
In the inner-most regions of radiation pressure supported accretion disks, 
the turbulent magnetic 
pressure may greatly exceed that of the gas.  If this is the case, it is 
possible for bulk Alfv\'enic motions driven by the magnetorotational
instability (MRI) to surpass the electron
thermal velocity.  Bulk rather than thermal 
Comptonization may then be the dominant radiative process which mediates
gravitational energy release.  For sufficiently large 
turbulent stresses, we show that turbulent Comptonization produces 
a significant contribution to the far-UV and X-ray emission of black hole
accretion disks. The existence of this spectral component 
provides a means of obtaining direct observational constraints on the nature 
of the turbulence itself.  We describe how this component may 
affect the spectral energy distributions and 
variability properties of X-ray binaries and active galactic nuclei.

\end{abstract}


\section{Introduction}

The thin accretion disk is the model most commonly employed in explaining gravitational energy release in active galactic nuclei (AGN) and X-ray binaries.  In the original formulation (Shakura \& Sunyaev 1973, Novikov \& Thorne 1973 NT73), both the mechanical structure of the disk and the thermal spectrum which it radiates were solved together, self consistently.  More recently, the nature of the anomalous ``$\alpha$-viscosity'', required for angular momentum transport, has been revealed to be MHD turbulence driven by the magnetorotational instability (MRI, Balbus \& Hawley 1991).  Spectral models taking into account reprocessing of nonthermal X-ray photons, along with the implied reflection features (Lightman \& White 1988, Guilbert \& Rees 1988), have proven to be very useful tools in diagnosing the variability and geometrical properties of such flows.  Further, the idea that the  broad-band X-ray power-law continuum results from thermal Comptonization of soft seed disk photons by a hot diffuse corona (Haardt \& Maraschi 1991) qualitatively reproduces the observed X-ray continuum in both AGN and X-ray binaries.  

Despite these advances, progress towards our understanding of the mechanics and radiative processes of accretion flows have occurred separately with little quantitative connection between the two.  We propose that bulk Comptonization of soft thermal disk photons by the turbulent motions themselves, in other words turbulent Comptonization, may substantially alter the dynamics and the emitted spectrum concurrently.  Thus, turbulent Comptonization allows for a direct concrete link between the details of the disk dynamics and the observed spectrum. 

Turbulent Comptonization is not an entirely new idea.\linebreak  Zel'dovich, Illarionov, and Sunyaev (1972) derived 
a general formulation in which turbulent Comptonization could be applied 
within a cosmological context.  Thompson (1994) considered Alfv\'enic 
reconnection-limited turbulent Comptonization while calculating the spectral energy distribution (SED) of 
a $\gamma$-ray burst fireball model.   At the same time, Thompson conjectured that this effect may be responsible for producing the Comptonized power-law 
continuum in AGN and X-ray binaries.  Rather than in a corona, we consider here the possibility that turbulent Comptonization occurs above the effective photosphere, {\it within} the body of the disk itself. 

The organization of this paper is as follows.  In section 2 we outline the basic hydrodynamic principles governing turbulent Comptonization in accretion disks while defining and estimating important spectral parameters such as the turbulent wave temperature and $y$-parameter.  In section 3, we estimate a characteristic timescale for turbulent
 Comptonization from the Kompan'eets equation in terms of fundamental thin disk parameters.  We show the results of Monte Carlo calculations in section 4 in order to demonstrate how turbulent Comptonization alters the emergent disk spectrum.  We consider how turbulent Comptonization may observationally manifest itself in AGN and X-ray binaries in section 5 and we summarize our conclusions in section 6.

\section{Turbulent Wave Temperature and $y$-parameter}

In black hole accretion disks, the turbulent velocity of the flow may be larger than the thermal velocity of the electrons in the inner-most regions.  As a result, electrons moving at the turbulent velocity will be able to Compton up-scatter soft photons to higher energy, potentially deforming the observed spectrum.  Here, we parameterize the turbulent motions in terms of a wave temperature $T_w$, which should be compared to the local electron thermal temperature of the disk $T_e$ in order to see whether turbulent or thermal Comptonization determines the radiative transfer.  At the same time, we deduce an effective turbulent $y$-parameter, allowing us to estimate the contribution to the resulting non-thermal spectrum.        

\subsection{Initial Estimates}

MRI turbulence is fundamentally magnetic in behavior with a characteristic velocity given by the Alfv\'en speed $v_A$ which is less than the local sound speed $c_s$.  Close to the hole, where the radiation pressure is dominant, $c_s$ is given by the radiation sound speed, which is larger than the sound speed of the gas.  Thus, it is possible for the MRI turbulence to be {\it highly supersonic relative to the gas}.

We propose that supersonic MRI eddies in the inner radiation pressure dominated region of thin accretion disks Compton up-scatter soft thermal photons arising from viscous dissipation.  In order for this to take place, the electron bulk velocity which corresponds to the turbulent velocity must be greater than the local electron thermal velocity.  

To show this, it is convenient to define a turbulent wave temperature $T_w$  
\be
\frac{k_BT_w}{m_e}\equiv\frac{1}{3}\left<v^2_w\right>.
\label{wavetemp}
\ee
MRI turbulence is ``strong'' on the outer scale such that $v_w\simeq v_A$.  That is, the kinetic energy density of the turbulence is roughly equal to the magnetic energy density (Hawley, Gammie, \& Balbus 1996).  If $P_w$ is the magnetic wave pressure given by
\be
P_w=\left<\frac{B^2}{8\pi}\right>
\ee
we then have a means of relating the turbulent velocities of electrons to the total (gas+radiation) pressure of the disk.  Recalling the fundamental assumption of thin disks, namely the viscous stress $\tau_{r\phi}$ is related to the total pressure $P$ by the proportionality relation $\tau_{R\phi}\simeq\alpha P$, we may write for MRI turbulence
\be
\tau_{R\phi}= \left< \rho v_R v_{\phi}-\frac{B_R B_{\phi}}{4\pi} \right>=\alpha\,P.
\ee
That is, the viscous stress is the sum of both the spatially correlated Reynolds and Maxwell stress of the turbulence.  In every numerical simulation of accretion disk turbulence, the Maxwell stress $\left<{B_R B_{\phi}}/{4\pi}\right>$ exceeds the Reynolds stresses $\left<\rho v_R v_{\phi}\right>$ by at least a factor of a few.  Also as a result of shear, rotation, and its magnetic nature, accretion disk turbulence is highly anisotropic such that the components of the Alfv\'en velocity obey\linebreak $\left<v^2_{A\phi}\right> \geq\, \left<v^2_{AR}\right> \geq \left<v^2_{Az}\right>.$  This implies $\tau_{R\phi}\neq P_w$, though they are of comparable magnitudes.  To quantitatively compare $\tau_{R\phi}$ and $P_w$, we introduce a parameter of proportionality $\psi$ such that
\be
\left<\frac{B^2}{8\pi}\right>=P_w=\psi\tau_{R\phi}=\psi\left< \rho v_R v_{\phi}-\frac{B_R B_{\phi}}{4\pi} \right>.
\ee
$\psi$ is a detailed function of the turbulence itself, as it measures  the level of anisotropy of the flow.  Now, we may write the square of the Alfv\'en velocity as
\be
\left<v^2_A\right> &  \simeq & \frac{2P_w}{\rho}\simeq 2\,\alpha \,\psi \,\frac{P}{\rho}\nonumber\\
& \,\,\,\,\,\,\,\,\,\,\,\,\,\,{\rm or} & \nonumber\\
\left<v^2_w\right> &  \simeq  & \left<v^2_A\right> \simeq 2\,\alpha\,\psi\, c^2_s.
\ee  
where we have defined $P/\rho\equiv c^2_s$.  The turbulent wave temperature is tied to the sound speed and thus the Alfv\'en speed in much the same way as the gas temperature is tied to the electron thermal velocity.  In terms of thin disk parameters, the turbulent wave temperature may be written as
\be
T_w & \simeq &\alpha\,\psi\,\frac{3m_e c^2}{2 k_B}\,\frac{{\dot m}^2}{\epsilon^2}\,r^{-3}\,\frac{{\mathcal D}^2}{{\mathcal B}{\mathcal C}}\,{\rm K} \nonumber\\
T_w & \simeq & 9.0\times 10^{9}\,\alpha\,\psi \frac{{\dot m}^2}{\epsilon^2}\,r^{-3}\,\frac{{\mathcal D}^2}{{\mathcal B}{\mathcal C}} \,{\rm K}.  
\label{kerrwavetemp}
\ee
Here, ${\dot m}$ is the accretion rate in units of the Eddington accretion rate ${\dot M}_{\rm edd}$, $\epsilon$ is the radiative efficiency, $r$ is the cylindrical radius in units of gravitational radii $r_g=GM/c^2$, and ${\mathcal A}-{\mathcal D}$ are the general relativistic correction factors (Riffert \& Herold 1995).  

We emphasize that the wave temperature defined above describes only the outer scale of the turbulence.  Turbulent flows may possess power for many decades of wavelength, as is thought to be the case with MRI turbulence (Hawley, Gammie, \& Balbus 1996).  The general picture of MRI turbulence is that the MRI continuously injects energy into the flow at the outer scale $\lambda_0$ which is somewhat smaller than the local disk scale height $H$.  Smaller scale modes then remove energy from the MRI near the outer scale via mode couplings where the eddy turnover time is roughly the dynamical time on the outer scale given by the Keplerian orbital frequency $\Omega^{-1}$.  When considering a scale $\lambda$ smaller than the MRI stirring scale, mode amplitudes will most likely be reduced - a point which we shall consider later.                

The turbulent wave temperature is {\it independent of black hole mass} and approaches the electron rest mass value $m_ec^2/k_B\simeq 6\times 10^{9}\,$K near the hole for large values of ${\dot m},\,\alpha$, and for $\psi\sim 1$.  Clearly, we see that the RMS bulk motion of the eddies can be much larger than the RMS electron thermal motion.  For a black hole mass of $M \sim 10^1$ and $10^8M_{\odot} $, the gas temperature at the inner-edge of the disk is $ T_e \sim 10^7$ and $10^5\,{\rm K}$, respectively.   The turbulent wave temperature, which is independent of black hole mass, can be as high as $T_w\sim 10^8-10^9\,{\rm K}$.  This is not surprising since the turbulent stresses, which scale like $\alpha P$, may be up to $10^{6}-10^7$ times greater than the gas pressure at the inner edge of black hole accretion disks.  Clearly, while modeling a thin $\alpha-$disk the effects of bulk electron motion due to vigorous MRI turbulence cannot be ignored in terms of spectral formation.  That is, the contribution to the kinetic energy of the electrons originating from random turbulent motions  can be much larger than the contribution given by random thermal motions from which a roughly thermal black body spectrum is presumably formed .  Notice that $T_w \propto  r^{-3}$ while the thermal black body temperature $\sim T_{{\rm th}}\propto r^{-3/4}$.  Thus, not only will $T_w$ be greatest in magnitude close to the inner edge, but the ratio  $T_w/T_{\rm th}$ will also be largest near the hole.  Therefore, effects due to large turbulent velocities and wave temperatures will operate most dramatically deep within the gravitational potential of the hole, where a large fraction of the gravitational power is released.  Also, disks which extend inward closer to the hole possess higher values of $T_w$, leading one to believe that disks around rapidly spinning Kerr holes are more prone to turbulent Comptonization than say, a disk surrounding a Schwarzschild hole.  

\subsection{Effects Due to Finite Mean Free Path: Optically Thick Reduction}

  If the local photon mean free path $\lambda_p$ is smaller than an eddy with characteristic size $\lambda$, the photons are effectively trapped by the eddies on that scale.  Thus, they will be  advected back and forth with the flow, not being able  to ``sample'' the turbulent velocities on the scale $\lambda$.  However, if we presume that MRI turbulence possesses power down to the scale of the photon mean free path $\lambda_p$, then the radiation may interact with the turbulent eddies on scales $\lambda\leq\lambda_p$, whose velocity amplitudes will be reduced with respect to those on the outer scale $\lambda_0$ by some amount.  This situation is similar to the case of acoustic radiation in stars where it is approximated that sound waves only interact with convective eddies of similar dimension (Goldreich \& Keeley 1977). 

Consider in particular the example of isotropic Kolmogorov turbulence.  In this case, the power spectrum of the turbulence $E(\lambda)\propto\lambda^{5/3}$, which implies 
\be
v_w({\lambda})\simeq v_w({\lambda_0})\left(\frac{\lambda}{\lambda_0}\right)^n
\ee   
where $n=1/3$.  To calculate the turbulent wave temperature $T_w$ on the scale of a the photon mean free path, we make note that $T_w\propto v^2_w$ giving us 
\be
T_w(\lambda_p)\simeq T_w(\lambda_0)\left(\frac{\lambda_p}{\lambda_0}\right)^{2/3}
\ee 
where $T_w(\lambda_0)$ is the wave temperature of the outer scale of the turbulence given by eq. (\ref{kerrwavetemp}).

As previously mentioned, accretion disk MRI turbulence is anisotropic.  In fact, MRI turbulence is {\it completely} anisotropic such that an eddy of a given velocity amplitude $v_{\lambda}$ possesses a spatial extent $(\lambda_R,\lambda_{\phi}, \lambda_z)$ where $\lambda_R\neq\lambda_{\phi}\neq\lambda_z$.  In order to proceed without a theory of accretion disk turbulence, we assume that an inertial range for the turbulence exists such that the properties of the turbulence possess a scale-free behavior.  This allows us to parameterize the velocity amplitude by
\be
v_w({\lambda})=v\left(\lambda_R,\lambda_{\phi},\lambda_z \right)\simeq v_A\left(\frac{\lambda_R}{L_R}\right)^r\left(\frac{\lambda_{\phi}}{L_{\phi}}\right)^s\left(\frac{\lambda_{\phi}}{L_{z}}\right)^t,
\ee
with $L_i$ being the dimension of the outer scale eddies in the $i^{\rm th}$ direction.  In the isotropic limit, $L_R=L_{\phi}=L_z$ and $r=s=t$.  We must note that accretion disks are Thomson optically thick and the diffusion approximation for the radiation transfer is reasonably accurate within the disk.  Thus, a photon scatters $\tau^2$ times before exiting the disk.  Since the photon field is roughly isotropic, we assume a given photon random walks a distance $\lambda_p$ in each direction at a given depth.  As a result, photons will interact with eddies whose velocities are given by
\be
v_w({\lambda_p})\simeq v_A\left(\frac{\lambda_p}{L_R}\right)^r\left(\frac{\lambda_p}{L_{\phi}}\right)^s\left(\frac{\lambda_p}{L_{z}}\right)^t=v_A\frac{\lambda^{\left(r+s+t\right)}_p}{L^r_R\,L^s_{\phi}\,L^t_z}.
\ee  
We will find it convenient to define 
\be
\eta_R\equiv L_R/L_z\,\,\,\,\,{\rm and}\,\,\,\,\,\eta_{\phi}\equiv L_{\phi}/L_z
\ee
which allows us to write
\be
v_w({\lambda_p})\simeq \frac{v_A}{\eta^r_R\eta^s_{\phi}}\left(\frac{\lambda_p}{L_z}\right)^n
\ee
where $\eta_{R,\phi}\geq 1$ for accretion disk MRI turbulence and $n=r+s+t$.  The corresponding wave temperature evaluated at the photon mean free path is then given by
\be
T_w\left(\lambda_p\right)\simeq \frac{T_w\left(\lambda_0\right)}{\eta^{2r}_R\,\eta^{2s}_{\phi}}\left(\frac{\lambda_p}{L_z}\right)^{2n}.
\ee

\subsection{Vertical Structure and $y-$Parameter}

The vertical structure of accretion disks is one of the biggest uncertainties in accretion disk theory.  In order to accurately model turbulent Comptonization in thin disks, the vertical structure must be accounted for.  If a soft photon is produced at some optical depth $\tau_0$ within the disk, it samples eddies of different scales as it climbs out of the disk, since its mean free path changes with density and thus with height.  We know neither the velocity amplitude of the turbulence nor the density profile as a function of depth for accretion disks.  In order to move forward, we must utilize an {\it ad hoc} prescription that takes the vertical structure into account in a manner suitable for modeling turbulent Comptonization while not over-constraining the disk atmosphere in any particular way.

One way of dealing with this uncertainty is to vertically average the turbulent stresses and the radiative transfer.  That is, we use the vertically averaged $\lambda_p$ while evaluating the wave temperature at all depths in the disk. By doing so, we relate $\lambda_p$ to the total optical depth $\tau$ since $\lambda_p\simeq (\kappa_{es}\rho)^{-1}$ and  $\tau\simeq H/\lambda_p$.  Effectively, we are modeling the disk to have a vertically constant density and turbulent stress.

We may relate the outer vertical scale of the turbulence $L_z$ to the height of the disk $H$ by introducing yet another parameter 
\be
L_z = \eta_z\, H
\ee 
where $\eta_z< 1$.  Recalling that $H/\lambda_p \simeq \tau$, the wave temperature becomes
\be
T_w(\lambda_p)\simeq\frac{T_w\left(\lambda_0\right)}{\eta^{2r}_R\,\eta^{2s}_{\phi}\,\eta^{2n}_z}\,\tau^{-2n}.
\label{reducedtemp}
\ee

\begin{figure}[b]
\plotone{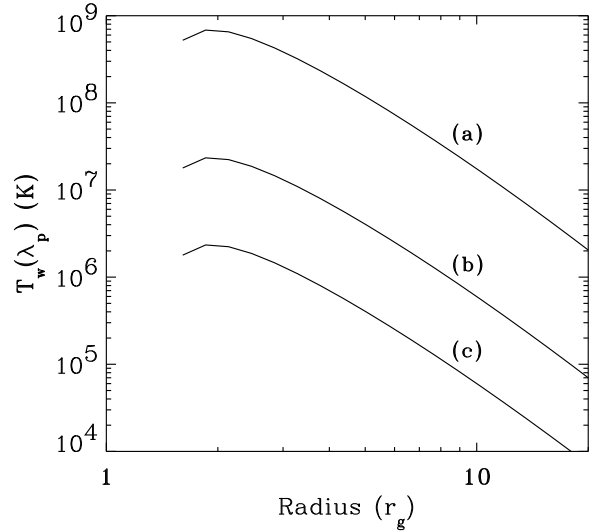}
\caption{
\label{fig:wavetemp} 
Turbulent wave temperatures evaluated at the scale $\lambda_p$ using the relativistic disk model of NT73 while including Riffert \& Herold's (1996) corrections to the vertical hydrostatic balance.  The black holes are maximally rotating with a=0.998. Curves (a) and (b) correspond to the disk parameters (${\dot m},\alpha$) equal to (1.0, 1.0) and (1.0, 0.1), respectively.  The cascade parameter $\psi=2.0$ and $n=1/3$ for all cases with $\eta^{2r}_R\eta^{2s}_{\phi}\eta^{2n}_z = 0.2$ for (b) and (c) while $\eta^{2r}_R\eta^{2s}_{\phi}\eta^{2n}_z =0.4$ for curve (a).  Curve (c) represents the case where (${\dot m}, \alpha$) equals ($0.1, 0.1$) with the turbulence confined to the upper layers with  an optical thickness of $\tau/10$ .}
\end{figure}
\begin{figure}[b]
\plotone{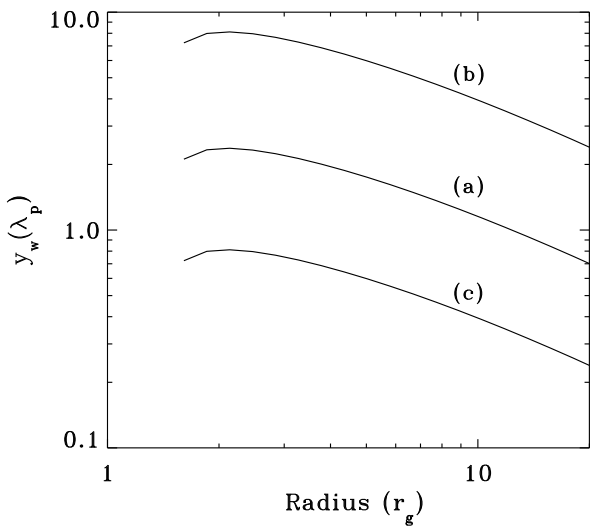}
\caption{
\label{fig:wavey}
Turbulent $y$-parameters for relativistic accretion disk models.  The labels (a)-(c) correspond to the curves in Figure (\ref{fig:wavetemp}) with the same labels.}
\end{figure}

Now, we may write the turbulent $y$-parameter as
\be
y_w\left(\lambda_p\right) & \simeq & \frac{4k_BT_w\left(\lambda_p\right)}{m_ec^2}\,\tau^2\nonumber\\
y_w\left(\lambda_p\right) & \simeq &\frac{4k_BT_w\left(\lambda_0\right)}{m_ec^2}\,\frac{\tau^{2\left(1-n\right)}}{\eta^{2r}_R\,\eta^{2s}_{\phi}\,\eta^{2n}_z}
\ee
where $\eta^{2r}_R\,\eta^{2s}_{\phi}\,\eta^{2n}_z\leq 1 $ and  $\tau^2$ was chosen since $\tau > 1$ for thin disks.  

Interestingly, eq. (\ref{reducedtemp}) tells us that $T_w(\lambda_p)\rightarrow T_w(\lambda_0)$ if
\be
\frac{\tau^{-2n}}{\eta^{2r}_R\,\eta^{2s}_{\phi}\,\eta^{2n}_z}\simeq 1.
\ee
This is only possible near the inner-edge since 
\be
\tau\simeq \frac{4\pi}{3\alpha}\frac{\epsilon\,r^{3/2}}{{\dot m}}\,\frac{{\mathcal B}{\mathcal C}}{{\mathcal A}{\mathcal D}}
\label{tau}
\ee
for thin disks.  Close to the hole, where a large fraction of the disk's power is released, $\lambda_p$ approaches $\lambda_0$ as long as ${\dot m}$ and $\alpha$ are sufficiently large.  When this is the case, $y_w$ is roughly given by the product of the wave temperature on the outer scale $T_w(\lambda_0)$ and the square of the optical depth 
\be
y_w\sim \frac{4k_BT_w(\lambda_0)}{m_e\,c^2}\,\tau^2.
\ee 
Equations (\ref{kerrwavetemp}) and (\ref{tau}) then allow us to write
\be
y_w\sim \frac{\psi}{\alpha}\left(\frac{4\pi}{3}\right)^2\,\frac{{\mathcal B C}}{{\mathcal A}^2}.
\label{youter} 
\ee
That is, when $\lambda_p\sim\lambda_0$, the turbulent $y$-parameter is approximately {\it independent of mass and radius}.

Figure \ref{fig:wavetemp} shows turbulent wave temperatures for maximally spinning black holes of varying ${\dot m}$ and $\alpha$.  The radius at which $T_w$ reaches its maximum value coincides with the most luminous radius of the disk, a consequence of $T_w$ directly scaling with the accretion stress.  The curves shown in Figure \ref{fig:wavetemp} are independent of black hole mass, a result of $c_s$ and $\tau$ being independent of black hole mass.  The curve labeled ``c'' depicts turbulent wave temperatures for ${\dot m}$ and $\alpha$ = 0.1.  For this particular case, the turbulence was confined to the upper-most 1/10 of the disk.  Confinement of the turbulence to a thin upper layer may be possible since regions of large magnetic pressure will be relatively buoyant.  Identical wave temperatures and $y$-parameters may be attained for a given ${\dot m}$ by reducing $\alpha$ by a factor $f$ while confining the turbulence to a narrow upper layer whose optical thickness is $\tau/f$, implying that the local turbulent stresses are $f$ times larger than the vertically averaged value.  Of course, this only applies for our constant density disk atmosphere model where each optical depth occupies an equal amount of height.
  
The turbulent $y$-parameters corresponding to the wave temperatures of figure \ref{fig:wavetemp} are shown in Figure \ref{fig:wavey}.  For our choice of $n$, $y_w\propto 1/r$ for radii substantially far away from the inner edge.  Yet, close to the inner edge, and in particular where $T_w$ and $y_w$ find their largest values, the dependence of $y_w$ on radius is weak which agrees with the estimate given by eq. (\ref{youter}).

\section{Compton Cooling of MRI Turbulence}

If soft seed photons created by viscous dissipation with energy $h\nu\sim k_B T_e$ are produced within the disk, they have a chance to ``resonantly'' interact with electrons belonging to eddies of spatial scale $\lambda\sim\lambda_p$.  If we assume that the radiation field is roughly isotropic, its evolution may be given by the Kompan'eets equation (1957) generalized for bulk motion (Hu, Scott, \& Silk 1994, Thompson 1994, and Psaltis \& Lamb 1997) 
\be
\frac{\partial n_{\nu}}{\partial t_c} & = & \left(\frac{k_BT_e}{m_e c^2}+\frac{\left<v^2_w\right>}{3\,c^2}\right)\left[4\nu\frac{\partial n_{\nu}}{\partial\nu}+\nu^2\frac{\partial^2n_{\nu}}{\partial^2\nu} \right]\nonumber\\
& \, &+\frac{h\nu}{m_ec^2}\left[4n_{\nu}+\nu\frac{\partial n_{\nu}}{\partial\nu}\right].
\label{Kompaneets}
\ee
$n_{\nu}$ is the photon occupation number and $t_c$ is the dimensionless Compton time where $d\,t_c \equiv n_e\sigma_Tcdt $.  Terms only quadratic in the turbulent velocity are retained since first order contributions vanish when a turbulent eddy is averaged over the domain of the flow.  It is clear that bulk electron motion produces spectral deformations in much the same way as thermal motion and when $\left<v^2_w\right>>3h\nu/m_e$ on the scale $\lambda_p$, photons stand to gain energy.

Equation (\ref{Kompaneets}) allows us to estimate a characteristic timescale $t_{c,w}$ for photons to thermally equilibrate with the turbulent motions.  
\be
\frac{\partial}{\partial t}\sim t^{-1}_{c,w}\sim \kappa_{es}\rho c\,\frac{\left<v^2_w\right>}{3\,c^2}=\kappa_{es}\rho c\frac{k_B T_w}{m_ec^2}. 
\ee  
$t^{-1}_{c,w}$ may also be thought of as the rate in which the eddies cool via Compton scattering.  In terms of physical disk quantities,
\be
t^{-1}_{c,w} & \sim & \frac{\kappa_{es}\rho}{c}\left<v^2_w\right>\,\sim \frac{\kappa_{es}\rho}{c}\frac{\alpha P}{\rho}\nonumber\\
t^{-1}_{c,w} & \sim & \frac{\tau}{c}\frac{\alpha P}{H\rho}\sim \frac{\alpha\,\tau}{c}\,\Omega^2 H.  
\ee  
Where $\Omega$ is the Keplerian orbital frequency. Further simplification is achieved by realizing that for thin disks 
\be
\frac{\alpha\,\Omega\,\tau H}{c}\sim{\mathcal{O}}(1),  
\ee
which is merely a statement that the disk removes thermal energy in the vertical direction by means of radiative diffusion at a rate given by $\sim\left(\alpha\Omega\right)^{-1}$.  We are left with
\be
t^{-1}_{c,w}\sim \frac{\alpha\,\tau}{c}\,\Omega^2 H\sim\Omega.  
\label{tcomp}
\ee
Note that in the expression above, only eddies on the outer scale were addressed and the cascade parameters $\psi$ and $\eta_i$ were discarded.  As mentioned in section 2.3, $\lambda_p$ approaches $\lambda_0$ near the inner-edge for sufficiently large values of ${\dot m}$ and $\alpha$.  When this is the case for unsaturated turbulent Comptonization, the photon spectral index $\Gamma$ and the turbulent wave temperature cut-off will vary with the turbulence on the dynamical time $\Omega^{-1}$ .  If the turbulent Comptonization is saturated, the photons will be able stay in thermal equilibrium with the turbulence, adjusting their own ``temperature'' to match that of the turbulence on the local dynamical time of the disk.

Turbulent Comptonization removes energy from the cascade most effectively on the scale $\lambda_p$, while directly placing that same amount of energy into the radiation field.  The mechanical energy flux which proceeds down to smaller wavelengths is diminished for scales $<\lambda_p$.

\begin{figure}[b]
\plotone{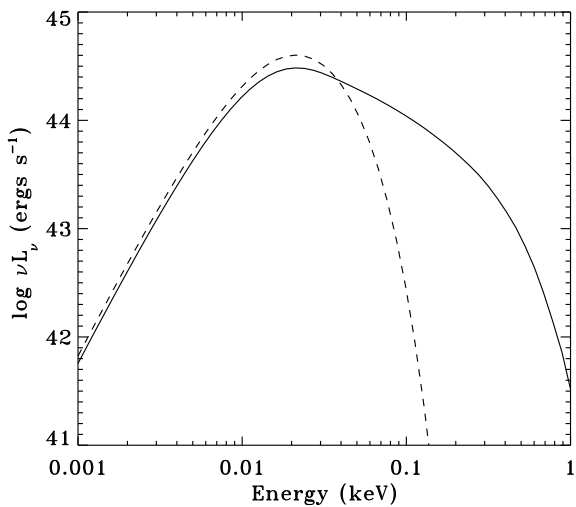}
\caption{
\label{fig:softxcess}
Accretion disk spectrum for a $10^8M_{\odot}$ black hole.  The turbulent Comptonization parameters where chosen from the curves labeled ``c'' in figs. (\ref{fig:wavetemp}) and (\ref{fig:wavey}). The corresponding multi-temperature disk blackbody is given by the dashed curve.}
\end{figure}

\section{Monte Carlo Calculation}

To estimate the effects of turbulent Comptonization on accretion 
disk
spectra, we use Monte Carlo radiative transfer calculations to generate 
models 
of integrated emission from the innermost radii.  The Monte Carlo 
calculations make 
use of the methods described in Pozdnyakov, Sobol', and Sunyaev (1983). We 
consider 
a maximally spinning  Kerr black hole with emission extending to the 
innermost stable 
circular orbit where the stress is assumed to vanish. Since much of the 
emission 
then comes from material moving at high velocities deep within the 
gravitational 
potential, we must also account for the effects of special and general 
relativity. 
To properly model the photon geodesics, we use a relativistic transfer 
code (Agol 1997).

We calculate the specific intensity emerging from $\sim 15$ logarithmically 
spaced 
annuli within $\sim 20\, r_g$ of the black hole. For each radius we model the 
turbulent region of the disk as a homogeneous plane-parallel slab. Photons drawn from a 
Planck 
distribution at the local disk effective temperature $T_{eff}$ are injected at the base of the slab and Compton 
scatter until they reach the surface. There is no emission or absorption 
within
the slab. For convenience, the scattering electron population is drawn from a
Maxwell-Boltzmann
distribution with temperature $T_w(\lambda_p)$.  MRI turbulence 
is
anisotropic such that electrons in an accretion disk will not be well 
represented by this distribution.  However if the radiation field is 
nearly
isotropic, the Kompan'eets equation (\ref{Kompaneets}) indicates that the effects of 
turbulent Comptonization
depend only on the magnitude of the electron velocity, not its 
direction.
The resulting spectra are not very sensitive to this assumption since the disk is optically thick, implying a highly isotropic photon field.  Similarly, any velocity distribution with the same second moment will likely produce a 
similar photon spectrum, so our choice of a Maxwell-Boltzmann distribution is
not restrictive. 
Further, the luminosity of a given annulus is normalized to the locally available gravitational power. 
To calculate the total spectrum we integrate over the first 20 $R_g$ using 
the 
relativistic transfer code to interpolate on our radial grid.  For radii 
where 
$T_{eff} > T_w(\lambda_p)$ we assume blackbody emission at $T_{eff}$.

Our Monte Carlo calculations are limited by 
the
uncertainties that plague vertical structure models of accretion disks. We neglect 
distributed emission and absorption and choose a Planck distribution with 
$T_{eff}$ 
for our soft input since it is difficult to determine how turbulent Comptonization will alter the vertical temperature and density profile of the disk.
In all cases we have calculated, the turbulent slab is effectively thin
at the photon frequencies of interest, at least at radii at which turbulent
Comptonization affects the spectrum, so our neglect of absorption is
justified.  Though our model of the disk atmosphere is crude, we believe
it gives a reasonable estimate of the magnitude and shape of the
resulting spectral deformation.

\begin{figure}[b]
\plotone{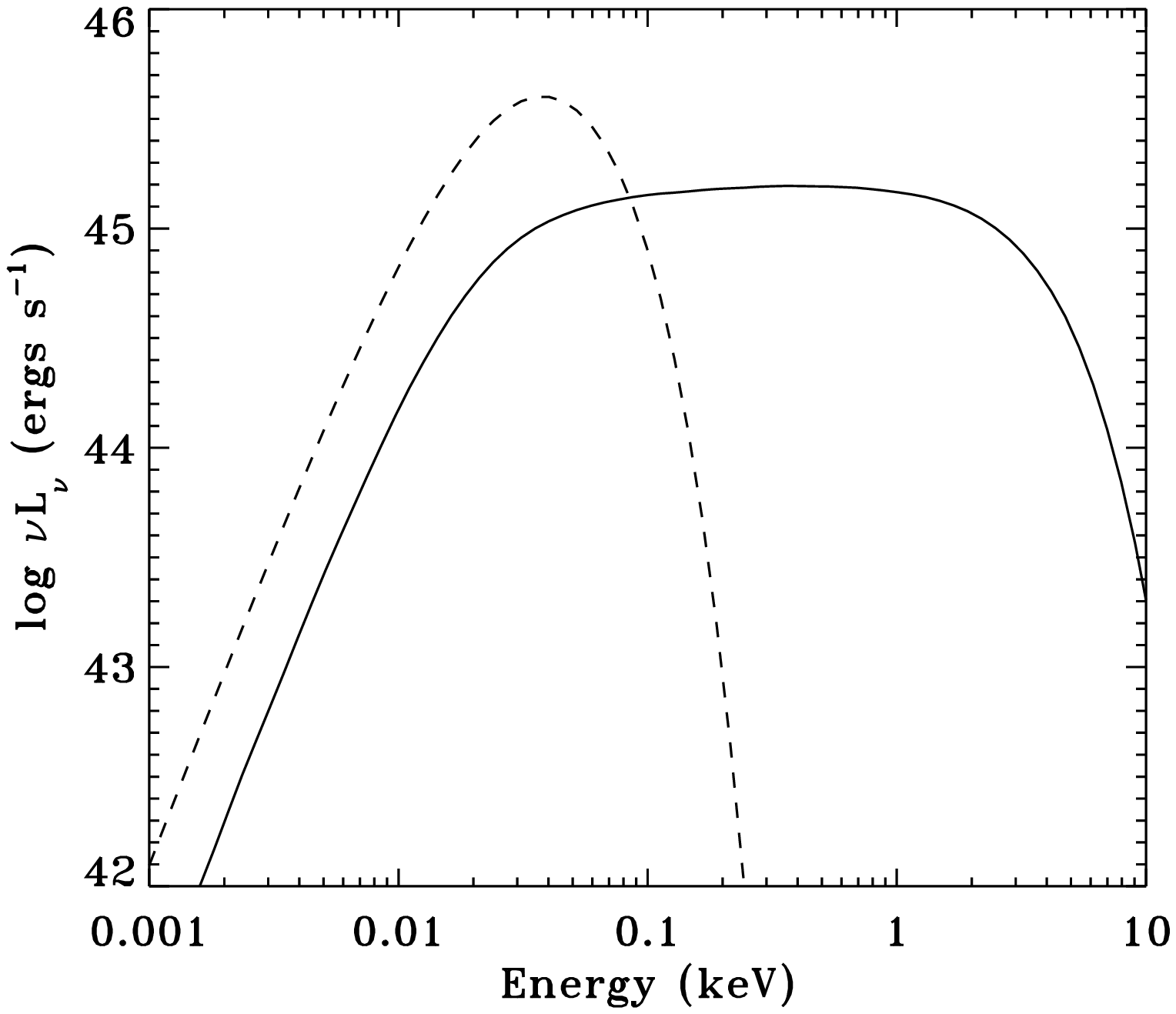}
\caption{
\label{fig:harderxcess} 
Accretion disk spectrum for a $10^8M_{\odot}$ black hole.  The turbulent Comptonization parameters where chosen from the curves labeled ``b'' in figs. (\ref{fig:wavetemp}) and (\ref{fig:wavey}). The corresponding multi-temperature disk black body is given by the dashed curve.\\}
\end{figure}

Figures \ref{fig:softxcess} and \ref{fig:harderxcess} show computed spectra for a $10^8M_{\odot}$ black hole with different radial profiles of wave temperature and $y$-parameter viewed from an inclination of $25^{\circ}$. Figure \ref{fig:softxcess} displays unsaturated turbulent Comptonization since $y_w<1$ for the entire disk while Figure \ref{fig:harderxcess} shows saturated turbulent Comptonization due to this particular disk model's large values of $y_w$.  Since $y_w$ is roughly uniform where most of the X-rays are formed via turbulent Comptonization i.e., near the inner-edge, the total X-ray spectrum has a well defined slope and spectral index.     

Standard thin disk models predict an effective optical depth $<1$ for large values of ${\dot m}$ and $\alpha$ near the inner edge.  As a result, the gas must
reach very high temperatures in order to produce sufficient photon flux to
balance the turbulent dissipation.  Turbulent Comptonization might alter this
picture, because it will itself amplify the available soft photon flux as
well as prevent energy from flowing down to the microscopic dissipation scale.
The gas temperature may therefore be cooler than would be expected in models
that only include bremsstrahlung and thermal Comptonization.

\section{Observational Implications}

If large turbulent stresses exist in radiation pressure supported accretion disks, then bulk Comptonization from the turbulent eddies may provide a large fraction of the far-UV and X-ray power in black hole or even neutron star sources.  The exact role turbulent Comptonization will play within a detailed spectral model is a difficult question to address.  The amount of power which escapes down to the dissipation scale $\lambda_D$ of the flow depends on the details of the turbulence on larger scales.  At the same time, the properties of the flow on relatively large scales are presumably determined by the vertical structure of the disk, which in turn is affected by the amount of mechanical energy being converted into thermal energy at $\lambda_D$.

In the last decade, paradigms for the broad-band continuum spectra for both AGN and X-ray binaries have begun to solidify.  Most spectral models of black hole accretion disks assume a hot diffuse ($\tau\leq 1$) population of thermal electrons above the disk (in other words, a corona) in order to reproduce the apparent Comptonized spectrum observed in the X-ray band.  Two-phase models of this sort (e.g. Haardt \& Maraschi 1991, 1993) reproduce the hard power law tail with the correct spectral index $\Gamma\sim 1.7$ with a Compton reflection hump as long as nearly all the gravitational power is locally released in the corona, perhaps due to buoyancy and magnetic reconnections of MRI turbulence (Haardt, 
Maraschi, \& Ghisellini 1994).  

        However, none of the current models take into account the details of the accretion flow in a predictive manner.  If turbulent Comptonization substantially contributes to the total output of the accretion flow, it then becomes possible to perform both spectral and hydrodynamic modeling concurrently since both the hydrodynamic and radiative processes are co-dependent.  In other words, the accretion disk spectrum directly measures both the flow, thermal structure, and geometry simultaneously.         

We now discuss how turbulent Comptonization may help explain some observations in both galactic and massive black hole sources.  Our purpose is not to offer definite conclusions, but to merely show how this effect may change (or complicate) the interpretation of accretion disk spectra.  

\subsection{Active Galactic Nuclei }

In many Seyfert 1 galaxies, a powerful extreme ultraviolet/soft X-ray feature exists above the extrapolated hard (2-10 keV) power law.  This component, known as the soft-excess (Walter \& Fink 1993), is commonly thought to arise from Comptonization with yet another thermal population of electrons resulting in a power law that is steeper than that of the 2-10 keV continuum.  Current disk-corona models cannot account for the soft excess in a robust fashion.  If the soft excess results from Comptonization, it is difficult to explain how and where a ``second'' corona comes about. A slim disk model (Abramowicz et al. 1988) has the ability to produce thermal soft X-rays, but the inclination angle of the disk has to be restricted to small values.  An alternative way of explaining the soft-excess is by assuming that it originates from turbulent Comptonization.  For large accretion rates ${\dot m}\sim 1$ and turbulent stresses $\alpha\sim 0.1$, turbulent wave temperatures approach $T_w\sim 10^7\,$K.  Lowering ${\dot m}$ and/or $\alpha$ may yield similar wave temperatures if the turbulent stresses are contained within a relatively small number of optical depths.  Narrow Line Seyfert 1 galaxies (NLS1) tend to have significant soft excesses which are rapidly varying (Boller, Brandt, \& Fink 1996).  This implies that the production of soft X-rays takes place close to the hole for these presumably high Eddington sources.  Turbulent Comptonization may very well be the mechanism which produces the soft excesses in NLS1's since larger wave temperatures are possible for disks with high values of ${\dot m}$ while the majority of the power originates close to the hole where variability is expected to be most rapid.  For example, Figure \ref{fig:softxcess} depicts turbulent Comptonization for a $10^8M_{\odot}$ black hole accreting at a rate ${\dot m}=0.1$, typical of NLS1 sources.  The spectral index $\Gamma$ is greater than two and the thermal cut-off is $\sim 1$keV.

\begin{figure}
\plotone{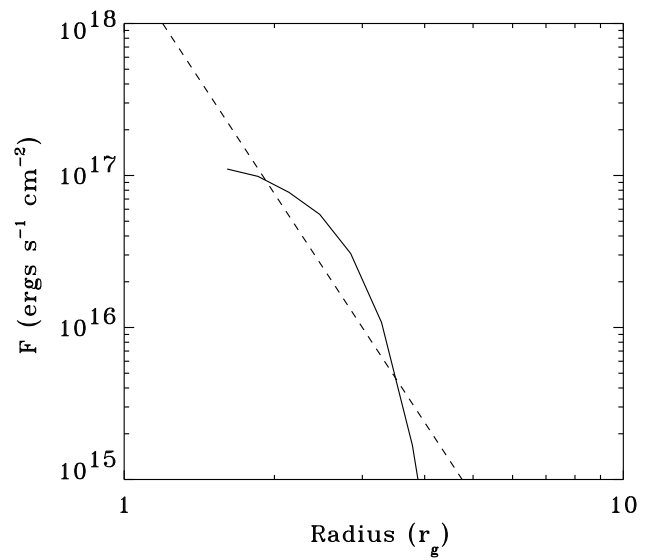}
\caption{
\label{fig:Iron} 
Flux in the neutral FeK edge band (7.0-7.5 keV) taken from the disk spectra computed in fig. (\ref{fig:harderxcess}).  The dashed curve corresponds to a disk flux $\propto r^{-5}$.     }
\end{figure}

 Studying the variability properties of NGC 3516, Edelson et al. (2000) sampled the optical, UV, soft, and hard X-ray bands continuously for nearly three days.  They found that the modulations of the non-thermal X-ray flux and optical/UV thermal flux where not correlated.  Either up-scattering or reflection models require soft thermal photons originating from the optically thick disk to either lead or follow variations in the hard optically thin component, which apparently is not the case for NGC 3516.  However, this may be reconciled if a special, source-dependent, geometry is invoked (Zdziarski, Lubi\'nski, \& Smith 1999 and Chiang \& Blaes 2001 in particular, for the case of NGC 3516) or if the corona consists of separate discrete blobs moving at mildly relativistic speeds away from the disk (Beloborodov 1999).                   

A different way of explaining the broad-band variability of NGC 3516 is by assuming the X-rays are produced in the inner-portions of the disk by turbulent Comptonization.  As long as the disk is thin, the hard output will not substantially interact with the cooler outer parts of the disk thus temporally disconnecting the hard from the soft output.  If the turbulent $y$- parameter is large, then very little UV/optical output emanates from the region of the disk that yields hard photons due to the large Compton amplification.  On the other hand, if $y_w\lesssim 1$, then UV output which corresponds to the inner X-ray region will vary on the thermal timescale $t_{\rm th}\sim (\alpha\Omega)^{-1}$.  From eq. (\ref{tcomp}), we see that the X-ray power in a given band can undergo fluctuations on timescale of order $\Omega^{-1}$, which can be much shorter than $t_{\rm th}$ depending on the magnitude of $\alpha$ and the details of the MRI cascade.  Another possibility is that the soft output from the turbulent Comptonizing region may contribute relatively little to the optical/UV luminosity compared to regions of the disk of larger radius as a result of the smaller emitting area.  

 In the case of NGC 7469 (Nandra et al. 1998), the lack of correlation between the UV and 2-10 keV X-ray luminosity suggests that neither an up-scattering model nor a reflection model can explain the time lags and correlations (or lack thereof) between the X-ray and UV output.  More recently however, Nandra et al. (2000) modeled the {\it RXTE} spectral data of NGC 7469 and found that the UV power of the source was correlated with the inferred spectral index $\Gamma$, rather than the amplitude, of the 2-10 keV band.  This implies that the overall X-ray spectral feature became softer while the total X-ray luminosity increased.  Their interpretation of the UV-$\Gamma$ correlation was that an increase in the UV flux led to enhanced cooling of the hot X-ray source which is a direct outcome of Compton up-scattering models (see also Chiang 2002).  Alternatively, the steepening of the X-ray component, along with its increase in power, may result from turbulent Comptonization.  That is, changes in ${\dot m}$ proportionally lead to changes in both the thermal UV luminosity and $T_w$. For values of $T_w\sim 10^6-10^7\,$K, $\Gamma$ will roughly increase with ${\dot m}$ as will the soft X-ray luminosity.  Further, one would expect the thermal cutoff of the high energy X-ray power-law to remain approximately fixed if turbulent Comptonization is responsible for the enhanced soft X-ray power as along as there is no direct physical correlation between the the parameters of the turbulent Comptonizing region ($T_w$, $y_w$) and the electron temperature of the hot corona.  Coronal up-scattering models however, would predict the thermal cut-off to decrease in energy.  Future missions which have enough sensitivity to detect changes in the cut-off energy may therefore be able to distinguish between these models.

The Fe K$\alpha$ line emissivity of MGC-6-30-15 may be another observational signature of turbulent Comptonization.  A significant red tail in the line profile indicates a Kerr geometry (Iwasawa et al. 1996) where the effects of turbulent Comptonization are potentially at their greatest.  Wilms et al. (2001) fit the Fe K$\alpha$ line emissivity to a radially-dependent power law $\propto r^{-\beta}$ with $\beta\sim 4-5$.  To account for such a steep profile, they required a powerful centrally located hard X-ray source with incident flux irradiating only the very innermost regions of the disk.  Those authors as well as others (Ballantyne, Vaughan, \& Fabian 2003) speculate that the steep emissivity profile may result from the extraction of spin energy via magnetic torques from the black hole itself (Blandford \& Znajek 1977).  Perhaps a milder way of interpreting the steep X-ray emissivity is by invoking turbulent Comptonization.  Figure \ref{fig:Iron} shows the radial dependence of the X-ray flux in the $7-7.5$ keV band due to turbulent Comptonization within a relativistic thin disk.  This is consistent with a power-law dependency of $\beta\sim 5$.  The eddies that have the ability to ionize the K-shell of iron are solely located close to the hole, near the marginally stable orbit.  This provides a natural explanation for concentrating energy release in the form of  X-rays near the central object.  If the Fe K$\alpha$ line emission {\it originates from within the optically thick} disk via turbulent Comptonization, then the ionization and thermal structure of the disk must be carefully modeled in order to determine whether or not this effect reproduces the observed equivalent widths.

\subsection{Galactic Black Holes}

Cyg X-1 is the most highly studied galactic black hole (GBH) source.  Like other GBH, it is observed in both the hard (low) and soft (high) spectral states (Done 2002) while mostly spending its time in the low state.  For GBH in general, the soft spectral state is dominated by a cool optically thick black body while a steep Comptonized power law extends deep into the hard X-ray/soft $\gamma$-ray region of the spectrum.  In the hard state, the photon output is dominated by a somewhat flat hard Comptonized power law with a moderate thermal contribution from the cool optically thick disk.  The measured luminosities of GBH in the hard state are typically lower than the soft state, implying a smaller ${\dot m}$ and/or smaller radiative efficiency compared to sources in the soft state.                 

Gierli\'nski et al. (1997, hereafter G97) found that in the low state,  the SED of Cyg X-1 could not be modeled with a standard disk-corona geometry.  In order to fit the observed spectra, they required a hot 50 keV optically thick ($\tau\sim 6$) Wien-like spectral component in addition to cool thermal disk emission and a hard relatively optically thin ($\tau\sim 1-2$) 100 keV power law, presumably from thermal Comptonization.  They concluded that the observed photon-starved spectrum may be explained by a geometry similar to the one proposed by Shapiro, Lightman, \& Eardley (1976) i.e., a geometrically thick hot central corona surrounded by a cool optically thick disk.        

In a later study, Gierli\'nski et al. (1999, hereafter G99) modeled Cyg X-1 in the soft state during a hard to soft state transition (Zhang et al. 1997).  The hard power law extended to $\sim 600$ keV with a spectral index $\Gamma\sim 2.5$ implying a very optically thin plasma with $\tau\sim 10^{-2}$.  If thermal Comptonization were responsible for the hard continuum, bumps would appear in the X-ray spectrum due to the distinct scattering orders. However, these features are not present.  As a solution, G99 proposed that the Comptonizing corona was composed of a hybrid thermal/non thermal plasma whose distribution function was roughly the sum of a relatively cool $\sim 50$ keV Maxwellian and a power law.      

Zhang et al. (1997) noted that during the state transition, the bolometric luminosity of Cyg X-1  changed very little.  From this fact, Poutanen, Krolik, and Ryde (1997) PKR97 put forth the proposition that the observed state transition of Cyg X-1 corresponded to a change in the mechanical state of the disk characterized by the inner edge moving inward as the spectral state changed from hard to soft.  This claim was supported by the fact that the temperature of the thermal disk component increased as Cyg X-1 evolved into the soft state.  The increase in the number of soft photons then has the ability to cool the corona, thus steepening the hard continuum.

Adopting this general picture, further details regarding the accretion flow of Cyg X-1 can be inferred if the $\sim 50\,$keV spectral component fitted by G97 and G99 is assumed to result from turbulent Comptonization.  If we approximate that the bolometric luminosity $L=\epsilon {\dot M}c^2$ and $T_w$ are constants during the state transition, then $y_w$ is the relevant parameter which is allowed to change.  In the hard state, the 50 keV component was fit with a Wien-law, implying  saturated Comptonization while in the soft state, the 50 keV component was fit with a steep power law, typical of unsaturated Comptonization.  Thus, during the transition from the hard to soft spectral state, $y_w$ decreased.  As previously discussed, the outer scale of the turbulence $\lambda_0$ may approach the photon mean free path $\lambda_p$ near the inner edge of the disk, where the spectral contribution from turbulent Comptonization is greatest.  With this, eq. (\ref{youter}) shows us that $y_w\propto\alpha^{-1}$.  Thus, {\it $\alpha$ increased as $y_w$ decreased during the hard to soft transition}.  This is consistent with the constant luminosity and wave temperature condition by noting the dependencies of the wave temperature on the outer scale given by  eq. (\ref{kerrwavetemp}).  That is, the quantity ${\dot m}^2/\epsilon^2 r^{3}$ must decrease during the transition. A constant luminosity requires ${\dot m}\propto\epsilon^{-1}$ which constrains $1/\epsilon^4r^3$ to decrease.  But, $\epsilon\propto r^{-1}$ yielding the result that $r$ must have decreased, which is consistent with the hypothesis of PKR97.

To summarize, by making the assumption that the 50 keV component in both the hard and soft spectral state of Cyg X-1 is due to turbulent Comptonization we may then determine further details of the accretion flow.  If the proposition of PKR97 is utilized to explain the spectral evolution of Cyg X-1, the basic model dependencies of turbulent Comptonization necessarily require that $\alpha$ increased while ${\dot m}$ decreased during the spectral transition.

\section{Summary and Conclusions}

Turbulent Comptonization may prove to be an important mediator of  gravitational energy release in accretion disks.  If so, simultaneously modeling both the mechanics and radiative processes of the flow not only becomes possible, but {\it necessary} in order to accurately fit the data.  We have shown that the amplitude and shape of the spectral feature resulting from turbulent Comptonization coincides with observed components in both AGN and X-ray binaries.  If turbulent Comptonization does not manifest itself spectrally, then its absence in the SED of accretion flows constrains values of fundamental disk parameters such as $\alpha, {\dot m},\, {\rm and}\,\epsilon$.   

We emphasize that throughout this paper, we have assumed that turbulent accretion stresses may greatly exceed the gas pressure.  It is this fact that allows turbulent Comptonization to operate as an important process.  It has been argued that in radiation pressure supported disks, buoyancy limits the turbulent magnetic pressure to values below that of the gas (Sakimoto \& Coroniti 1981, 1989; Stella \& Rosner 1984). Further, we assumed that large turbulent stresses exist down to the scale of the photon mean free path.  However, there may be other dissipation mechanisms acting on scales larger than $\lambda_p$.  One such example is Silk damping of compressive motions in the turbulence (Agol \& Krolik 1998).  Recent numerical simulations of un-stratified radiation pressure supported disks show that Silk damping may be responsible for a significant (but not dominant) fraction of energy release (Turner et al. 2003).  Such simulations performed in a vertically stratified medium will address both Silk damping and buoyancy effects simultaneously.  

Another uncertainty in determining the relative importance of turbulent Comptonization is the vertical structure of the disk. The presence of turbulent Comptonization provides a direct\linebreak means of turning turbulent mechanical energy into photon power without requiring a cascade down to a microscopic dissipation scale.  How this affects the structure and dynamics of the flow in a time-averaged fashion is difficult to predict.  In terms of making detailed comparisons between theory and observations, these questions must be answered before any quantitative progress can be achieved.

\acknowledgements{The authors would like to thank Phil Arras, Philip Chang, and Neal Turner for many helpful conversations.  This work was supported by NASA grant NAGS-13228 and UCSB/LANL CARE grant SBB-001A.}

\end{document}